\documentclass[12pt,preprint]{aastex}

\newcommand{\be}{\begin{equation}}
\newcommand{\ee}{\end{equation}}

\newcommand{\km}{{\, \rm km}}
\def\refnew#1{(\ref{#1})}

\newcommand{\cm}{{\, \rm cm}}
\newcommand{\s}{{\, \rm s}}
\newcommand{\dyne}{{\, \rm dyne}}
\newcommand{\yr}{{\, \rm y}}
\newcommand{\gm}{{\, \rm g}}

\begin{document}

\title{Tidal Evolution of Rubble Piles}

\author{Peter Goldreich$^{1,2}$ and Re'em Sari$^{2,3}$}
\affil{$^1$Institute for Advanced Study, Princeton, NJ
\\$^2$Caltech 130-33, Pasadena, CA 91125
\\ $^3$Racah Institute of Physics, Hebrew University, Jerusalem 91904, Israel.}

\begin{abstract}
Many small bodies in the solar system are believed to be rubble
piles, a collection of smaller elements separated by voids. We
propose a model for the structure of a self-gravitating rubble pile.
Static friction prevents its elements from sliding relative to each
other. Stresses are concentrated around points of contact between
individual elements. The effective dimensionless rigidity,
$\tilde\mu_{rubble}$, is related to that of a monolithic body of
similar composition and size, $\tilde\mu$ by $\tilde \mu_{rubble}
\sim \tilde \mu^{1/2} \epsilon_Y^{-1/2}$, where $\epsilon_Y \sim
10^{-2}$ is the yield strain. This represents a reduction in
effective rigidity below the maximum radius, $R_{max}\sim
[\mu\epsilon_Y/(G\rho^2)]^{1/2}\sim 10^3\km$, at which a rubble pile
can exist. Densities derived for binary near-Earth asteroids imply
that they are rubble piles. As a consequence, their tidal evolution
proceeds $10^3$ to $10^4$ times faster than it would if they were
monoliths. This accounts for both the sizes of their semimajor axes
and their small orbital eccentricities. We show that our model for 
the rigidity of rubble piles is compatible with laboratory experiment
in sand.
\end{abstract}


\section{Introduction}

Rubble piles are bodies composed of smaller elements separated by
voids. There is compelling evidence that at least some small solar
system bodies are rubble piles bound by gravity. Their telltale
signature is a mean density below that of their constituent
elements. Examples include: four icy satellites of Saturn, the
coorbitals, Janus and Epimetheus, and the F-ring shepherds,
Prometheus and Pandora \citep{JaF04,PTW+07}, the rocky main belt
asteroids, C-type 253 Mathilde \citep{VTH+97,YBD+97} and M-type 22
Kalliope \citep{MaB03}, and the binary near-Earth asteroid 1999 KW4
\citep{OMB+06}. The largest of these bodies, Janus and Kalliope,
have dimensions of order $100\km$. It is unclear whether larger
rubble piles exist or whether all smaller bodies are rubble piles.

Intuitively we expect a rubble pile to be weaker than a monolithic
body of the same composition. Thus tidal dissipation at a rate that
is more rapid than typical for a monolith is considered evidence for
a rubble pile \citep{MaB03}. The orbits of binary near-Earth
asteroids are prime examples; the sizes of their semimajor axes and
their low orbital eccentricities suggest that they are evolving tidally
at rates between $10^3$ and $10^4$ times faster than estimates for
monolithic bodies of similar size would predict \citep{WaR06a}. In
what follows, we provide a theoretical basis for estimating tidal
dissipation rates in rubble piles and show that it can account for
this large speedup of tidal evolution.

The plan of our paper is as follows. In \S \ref{sec:modulus}, we
formulate a quantitative theory for the effective rigidity of a
self-gravitating rubble pile and demonstrate that it is due to voids
rather than cracks. Limits on the sizes of rubble piles are derived
in \S \ref{sec:sizes}. In \S \ref{sec:tides}, we apply our
theory to the tidal evolution of binary near-Earth asteroids.

\section{Effective Elastic Modulus Of A Rubble Pile}
\label{sec:modulus}

We begin by reviewing the tidal response of a uniform body of
density $\rho$, rigidity $\mu$, and radius $R$. As is customary,
we define the dimensionless rigidity by
$\tilde\mu$;
\begin{equation}
\label{tildemu}
{\tilde\mu}\equiv \frac{19\mu}{2g\rho R}\, . \label{eq:mubar}
\end{equation}
Next we show that $\tilde\mu$ is the ratio of the fluid strain to the
elastic strain.\footnote{Arguments in this section are order of
magnitude only.}

We assume that the tidal force, $f$, is weak in comparison to the
cohesive force of the body's self gravity, $gM$, where $g\sim G\rho
R$.
If the body were fluid, $\mu=0$, it would suffer a strain
\begin{equation}
\label{epsilong}
\epsilon_g\sim \frac{f}{g\rho R^3}\, , \label{eq:epsfluid}
\end{equation}
whereas if it were elastic but lacked self-gravity, $g=0$, the
strain would be
\begin{equation}
\label{epsilonmu}
\epsilon_\mu\sim \frac{f}{\mu R^2}\, . \label{eq:epsmu}
\end{equation}
To order of magnitude, the ratio between expressions \refnew{epsilong}
and \refnew{epsilonmu} reproduces $\tilde\mu$ given by equation
\refnew{eq:mubar}.

How does the tidal response of a rubble pile differ from that of a
monolith? To answer this question, we investigate some simple
models.

\subsection{cracks do not matter}
\label{subsec:cracks}

Normal stresses are seamlessly transmitted across cracks, so a body's
response to weak tides is unaffected by cracks. Consider a body of
radius $R$ composed of cubical elements whose sides have length $r \ll
R$. The ratio of the weight of a single cube, $gM(r/R)^3$, to the
divergence of tidal stress acting on its volume, $f(r/R)^3$, is just
$gM/f$. Thus a coefficient of static friction larger than $f/gM$ would
suffice to prevent the cubes from slipping relative to each
other. Coefficients of static friction for rocks and dry ice are of
order unity, and $f/gM\sim (R/a)^3$ for an equal mass binary with
semi-major axis $a$.

\subsection{voids are key}\label{subsec:voids}

\subsubsection{uniform spheres}\label{subsubsec:spheres}

Next we consider a body of radius $R$ composed of identical spheres
of radius $r\ll R$. Its mean density ${\overline\rho}\approx
0.7\rho$. A typical cross section cuts $(R/r)^2$ small spheres each
of which transmits forces $F(r/R)^2$ to its neighbors, where
$F\sim gM+f$ includes both tidal forces and self gravity. In so doing,
each small sphere undergoes a linear distortion $\delta x$ and forms
contact surfaces with its neighbors of area $\delta x\, r$.  Within
$(\delta x\, r)^{1/2}$ of each contact, the strain is of order
$(\delta x/r)^{1/2}$ so
\begin{equation}
\label{eq:balance} F\left(r\over R\right)^2\sim\mu r^{1/2} \delta
x^{3/2}\, .
\end{equation}
The average strain is just $\delta x/r$, where from equation
\refnew{eq:balance}
\begin{equation}
\frac{\delta x}{r}\sim \left(F\over \mu R^2 \right)^{2/3}\, .
\label{eq:aveps}
\end{equation}
Most of this strain is due to the body's self-gravity. To isolate
the tidal strain, we expand $F^{2/3}$ in equation \refnew{eq:aveps}
around $F\sim gM$ to obtain
\begin{equation}
\epsilon\sim \frac{f}{\mu R^2}\left(\frac{\mu}{g\rho
R}\right)^{1/3}\, . \label{eq:eqssphere}
\end{equation}
Thus the effective dimensionless tidal rigidity of a body composed
of identical spheres is
\begin{equation}
\label{eq:muspheres} \tilde\mu_{spheres} \sim \left(\mu \over g\rho
R\right) ^{1/3} \sim \tilde \mu^{2/3}\, .
\end{equation}
This result is equivalent to that originally established by
\cite{DuM57}.

\subsubsection{irregular fragments \label{subsubsec:irregular}}

Natural rubble piles are likely to be composed of irregularly shaped
elements whose surfaces have local radii of curvature, $\hat r$,
that are much smaller than the elements' sizes, $r$. Compared to
rubble piles composes of spheres, contact areas would be reduced,
stress concentrations increased, and the effective rigidity lowered.
A simple modification of the derivation given in
\ref{subsubsec:spheres} suffices to evaluate the effective rigidity
of a rubble pile, $\tilde\mu_{rubble}$. Each element still transmits
its share of the total force. However, $\hat r$ must replace $r$ on
the right hand side of equation (\ref{eq:balance}). Thus now
\begin{equation}
\frac{\delta x}{r}\sim \left(F\over \mu R^2
\right)^{2/3}\left(r\over {\hat r}\right)^{1/3}\, .
\label{eq:avepsp}
\end{equation}
Continuing as before, we find that the average strain across the
rubble pile is increased by the factor $(r/{\hat r})^{1/3}$ with the
consequence that the effective rigidity now reads
\begin{equation}
\tilde \mu_{rubble}\sim\tilde \mu_{spheres}
\left(\hat r\over r\right)^{1/3}\sim\tilde \mu^{2/3}
\left(\hat r\over r\right)^{1/3}. \label{eq:effmuirreg}
\end{equation}
The sharper the contact points, the softer the rubble pile, up to a
limit at which the stress near the contact surfaces reaches the
material's yield stress $\sigma_Y$, or yield strain
$\epsilon_Y=\sigma_Y/\mu$.  This limit is met at
\begin{equation}
\frac{\hat r}{r}\sim\frac{1}{\left(
{\tilde\mu\epsilon_Y^3}\right)^{1/2}}
\label{eq:minrhat}
\end{equation}
Sharper contact points than allowed by equation \refnew{eq:minrhat}
would be dulled by material flow or failure.  Therefore,
\begin{equation}
{\tilde \mu}_{rubble} \gtrsim\left({\tilde
\mu}\over \epsilon_Y\right)^{1/2}\, .
\label{eq:mintildemu}
\end{equation}
Experimentally it is generally found that the effective rigidity of
a granular material scales in direct proportion to the square root
of the confining pressure. \cite{God90} provides an explanation for
this scaling which is similar to ours.

Equations (\ref{eq:muspheres}) and \refnew{eq:effmuirreg}
demonstrate that the effective rigidity of a rubble pile is smaller
than that of a monolithic body of the same size. The reduction in
rigidity is independent of the sizes of the elements into which the
body is divided. It arises from the concentration of stresses due to
the presence of voids.

\begin{figure*}
\begin{center}
\includegraphics[angle=-90,width=16cm]{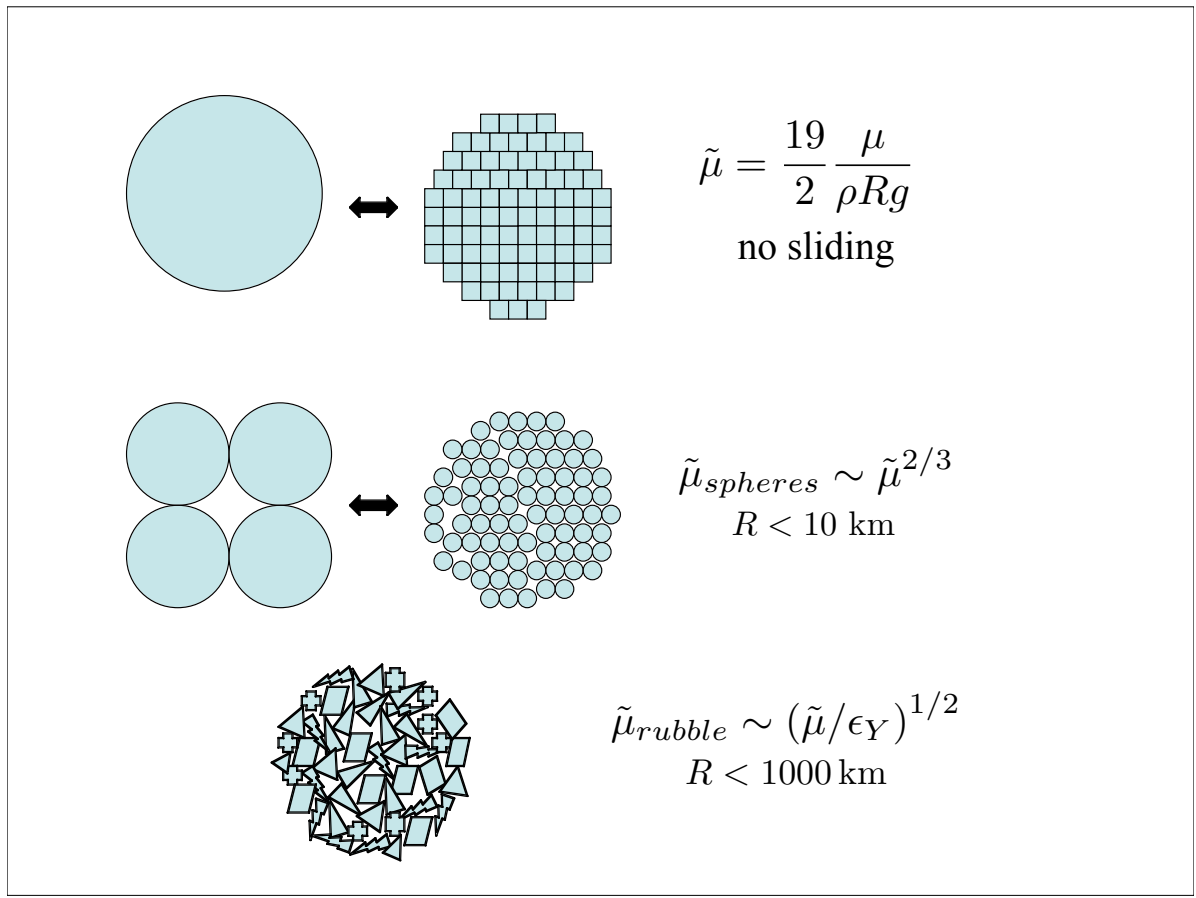}
\end{center}
\caption{Three simple models of fragmented bodies. Upper row depicts a
body composed of cubical elements. There are no voids. Static friction
prevents the elements from sliding relative to each other.  Its
effective rigidity is identical to that of a monolith. Middle row
shows a body composed of spherical elements. Voids are
present. Stresses concentrate near contacts between elements. The
effective rigidity is smaller than that of a monolith and is
independent of the sizes of the elements. Such an idealized
configuration requires the spheres to be sufficiently smooth. If made
of rock or ice, its radius could not be larger than about
$10\km$. Bottom row illustrates a more realistic rubble pile composed
of irregular elements. Sharper contact points increase stress
concentration more than for a body composed of spherical elements.
Accordingly, its effective rigidity is further decreased.  Radii of
rubble piles composed of rock or ice cannot be larger than about
$1000\km$.}
\label{fig:models}
\end{figure*}

\subsection{energy considerations}

We re-derive equation \refnew{eq:effmuirreg} based on energy
considerations. Strains of order $(\delta x/{\hat r})^{1/2}$ are
attained within a distance $(\delta x\, {\hat r})^{1/2}$ of the
contacts between individual elements. Thus the elastic energy stored
within the rubble pile satisfies
\begin{equation}
\delta E\sim \mu(\delta x)^{5/2}{\hat r}^{1/2}\left(R\over
r\right)^3\, . \label{eq:delEdelx}
\end{equation}
We can also express $\delta E$ in terms of the effective
dimensionless rigidity, ${\tilde\mu}_{rubble}$ and the
average strain in the rubble pile, $\delta x/r$ as
\begin{equation}
\delta E\sim {\tilde\mu}_{rubble}g\rho R\left(\delta
x\over r\right)^2 R^3\, . \label{eq:delEmueff}
\end{equation}
Equating the expressions for $\delta E$ given in equations
\refnew{eq:delEdelx} and \refnew{eq:delEmueff}, we arrive at
\begin{equation}
{\tilde\mu}_{rubble}\sim
{\tilde\mu}\frac{{\delta x}^{1/2}{\hat r}^{1/2}}{r}\, .
\label{eq:mudimone}
\end{equation}
Finally, by using equation \refnew{eq:avepsp} to eliminate $\delta
x$, we recover equation \refnew{eq:effmuirreg}.

\section{Critical Sizes For Rubble Piles}
\label{sec:sizes}

\subsection{mechanical limits}

At
\begin{equation}
R=R_*\sim \left({\mu\epsilon_Y^3 \over \rho^2 G}\right)^{1/2}\, ,
\label{eq:R1}
\end{equation}
which corresponds to $\tilde\mu\sim \epsilon_Y^{-3}$,
$\tilde\mu_{rubble}\sim \tilde\mu_{spheres}\sim \epsilon_Y^{-2}$ and
$\hat r/r\sim 1$. For nominal values of $\mu_{rock}\approx 5\times
10^{11}\dyne\cm^{-2}$, $\mu_{ice}\approx 3\times
10^{10}\dyne\cm^{-2}$, $\epsilon_Y\sim 10^{-2}$, $R_*\sim 10\km$ for
rubble piles composed of either rock or ice. Moreover,
$\tilde\mu_{rubble}\sim 10^4$ as compared to $\tilde\mu\sim 10^6$
for a monolith of radius $R_*$. At
\begin{equation}
\label{R2} R_{max}=\left(\mu \epsilon_Y \over \rho^2
G\right)^{1/2}\, ,
\end{equation}
which corresponds to $\tilde\mu\sim \epsilon_Y^{-1}$ and $\hat
r/r\sim \epsilon_Y^{-2}$, the contact areas are comparable to the
surface areas of individual elements so
$\tilde\mu_{rubble}\sim\tilde\mu$. With nominal parameters,
$R_{max}\sim 10^3\km$ and $\tilde\mu_{rubble}\sim \tilde\mu\sim
10^2$.

A body with $R<R_*$ would avoid elastic failure if it were composed
of identical spheres. For $R>R_*$, elastic failure would occur at
points of contact among spheres. More generally, we would expect the
voids in rubble piles to occupy a decreasing fraction of the volume
with increasing $R$ up to $R=R_{max}$. At $R_{max}$, the average
interior pressure $g\rho R_{max}\sim \sigma_Y$, so voids could only
exist near the surface.

\subsection{thermal limits}

Rubble piles should be more common among smaller bodies because they
cool more rapidly than larger ones and therefore are less likely to
have been melted. Thermal diffusivities of rock and ice are of order
$10^{-2}\cm^2\s^{-1}$, which implies
\begin{equation}
t_{cool}\sim 3\times 10^{10}\left(R\over 10^3\km\right)^2\yr\, .
\label{eq:tcool}
\end{equation}
Even bodies as small as $R_*\sim 10\km$ might have been melted if
they formed early and were endowed with short lived radioactive
nuclides. On the other hand, bodies as large as $R_{max}\sim 3\times
10^2\km$ which were fragmented by collisions after the short lived
radioactive nuclides had decayed could have avoided melting.

\section{Implications For Tidal Evolution\label{sec:tides}}

Tides play crucial role in orbital and spin evolution of binaries.
Here we focus on the evolution after the secondary's spin has become
synchronous with the mean orbital angular velocity while the
primary's spin remains much faster. In this case, tides raised on
the primary cause both the semimajor axis, $a$, and the orbital
eccentricity, $e$, to grow while those raised on the secondary have
negligible effect on the semimajor axis and cause the eccentricity
to decay \citep{Gol63,GoS66}.  Relevant expressions for $e\ll 1$
are:
\begin{equation}
\label{semimajor}
{1\over a}{da \over dt}=3 {k_p \over Q_p} {M_s \over M_p } \left( R_p
\over a\right)^5 n\,
\end{equation}
and
\begin{equation}
{1 \over e}\frac{de}{dt}={57 \over 8} {k_p \over Q_p} {M_s \over
M_p} \left( R_p \over a \right)^5 n \, ,
\end{equation}
for tides raised on the primary, and
\begin{equation}
\label{damp} {1 \over e}\frac{de}{dt}=-{21 \over 2} {k_s \over Q_s}
{M_p \over M_s} \left( R_s \over a \right)^5 n \, .
\end{equation}
for tides raised on the secondary\footnote{Subscripts $p$ and $s$
denote primary and secondary. We adopt standard notations for tidal
Love number, $k$, and quality factor, $Q$ \citep{MuD00}. $Q^{-1}$ is
a stand-in for $\sin{2\delta}$, where $\delta$ is the tidal phase
lag.}

Tidal evolution rates depend on two parameters, $k$ and $Q$. The
estimation of $k$ for monoliths involves little uncertainty. For a
body of uniform density,
\begin{equation}
k=\frac{1.5}{1+{\tilde\mu}}\, . \label{eq:k}
\end{equation}
The estimation of $Q$ is more uncertain. Available evidence suggests
that $Q\sim 10^2$ for monolithic bodies \citep{GoS66}.

\subsection{Semimajor axis evolution in binary rubble piles}

Semimajor axis evolution is driven by the transfer of angular momentum
from the spin of the primary to the orbit. Below, we estimate
timescales for the semimajor axes of some well observed binary NEAs to
have evolved from much smaller initial values to their current ones. Integrating
equation \refnew{semimajor}, we obtain
\begin{equation}
\label{age} T={2 \over 39} {Q_p \over k_p} {M_p \over M_s} \left( a
\over R_p \right)^5 \frac{1}{n}
\end{equation}
We compare timescales for models
in which the bodies are assumed to be monolithic solids, fluids, and
rubble piles. We set $Q=100$ in each case.

As the entries in table \ref{table} demonstrate, the timescale for
semimajor axis evolution is measured in Gyrs for monoliths, years
for fluids, and Myrs for rubble piles. Only the latter is consistent
with estimates of $10\,$Myr for the dynamical life time of NEAs
\citep{GMF00}. Since it is plausible that the stress concentration
in rubble piles results in $Q<100$, the ages we estimate for rubble
piles should be viewed as upper limits.

\begin{center}
\begin{table}
\scriptsize
\begin{tabular}{l|c|c|c|c|c|c|c}
Asteroid                & Orbital       & Semimajor & Primary       & Secondary & Monolith  & Fluid     & Rubble Pile \\
name                & period (days) & axis (km) & diameter (km)& diameter (km)& age (Gyr)   & age (yr)      & age (Myr) \\ \hline
(66391) 1999 KW$_4$     & 0.73      & 2.5           & 1.2           & 0.4           & 15            & 37            & 7.5       \\
1999 HF$_1$         & 0.58      & 7.0           & 3.5           & 0.8           & 3.6           & 74            & 5.2       \\
(5381) Sekhmet      & 0.52      & 1.5           & 1.0           & 0.3           & 4.2           & 7.0           & 1.7        \\
(66063) 1998 RO$_1$ & 0.6           & 1.4           & 0.8           & 0.38      & 4.1           & 4.4           & 1.3        \\
1996 FG$_3$         & 0.67      & 2.6           & 1.5           & 0.47      & 4.3           & 16            & 2.7        \\
(88710) 2001 SL$_9$ & 0.68      & 1.4           & 0.8           & 0.22      & 24            & 26            & 7.8        \\
1994 AW$_1$         & 0.93      & 2.3           & 1         & 0.5           & 14            & 23            & 5.6        \\
2003 YT$_1$         & 1.2           & 2.7           &1          & 0.18      & 880       & 1500      & 360    \\
(35107) 1991 VH     & 1.4           & 3.2           & 1.2           & 0.44      & 74            & 180       & 36         \\
2000 DP$_{107}$     & 1.8           & 2.6           & 0.8           & 0.3           & 540       & 580       & 180    \\
(65803) Didymos     & 0.49      & 1.1           & 0.8           & 0.17      & 11            & 12            & 3.7        \\
(5407) 1992 AX      & 0.56      & 6.8           & 3.9           & 0.78      & 2.1           & 54            & 3.4        \\
(85938) 1999 DJ$_4$ & 0.74      & 0.8           & 0.4           & 0.17      & 55            & 15            & 9.0        \\
2000 UG$_{11}$      & 0.77      & 0.4           & 0.2           & 0.08      & 280       & 18            & 22         \\
(3671) Dionysus     & 1.2           & 3.8           & 1.5           & 0.3           & 190       & 720       & 120    \\
2002 CE$_{26}$      & 0.67      & 5.1           & 3         & 0.21      & 88            & 1300      & 110    \\
\end{tabular}

\caption{\label{table} Ages of NEA binaries based on assuming their
semimajor axes have evolved from much smaller initial values [eq.
\refnew{age}]. Comparison for monolithic ($k=3/2\tilde \mu$), fluid
($k=3/2$), and rubble pile ($k=3/2\tilde\mu_{rubble}$) strength for
primary. Binary parameters from compilation by \cite{WaR06a}. }
\end{table}
\end{center}

\subsection{Comparison with experiments in sand}

The effective rigidity of our model rubble pile, $\mu_{rubble}$, is
proportional to the square root of the confining pressure and
independent of the size of the individual elements. Laboratory
measurements of the shear velocity, $c_s=\sqrt{\mu/\rho}$, in sand
as a function of pressure provide a useful calibration. The data on
$c_s(p)$ plotted in figure 1 of \cite{God90} are replotted in our
figure 2. On the figure's upper boundary we display the radius of an
asteroid whose average internal pressure
\begin{equation}
P=(4\pi /15) G \rho^2 R^2 \cong 2.2 \times 10^{3} \left( R \over
1\km \right)^2 \, {\dyne\cm^{-2}}\, \label{eq:aveP}
\end{equation}
with $\rho\cong 2\gm\cm^{-3}$ corresponds to that given on the lower
boundary. The range of pressures covered in the experiments on sand
correspond to those inside asteroids with radii from 10-40$\km$. The
right-hand boundary of the figure shows the effective rigidity
corresponding to the shear velocity. It is well-fit by the solid
line which is derived from our expression for effective rigidity
with $\epsilon_Y \cong 0.17$. This should not be taken as evidence
that the yield stain of sand is $0.17$ since our formula is only
accurate to order of magnitude. However, it does suggest that the
ages we estimate in table \ref{table} may be a factor of a few too
large. The dashed line indicates the higher effective rigidity of a
body composed of uniform quartz spheres.

\begin{figure*}
\begin{center}
\includegraphics[width=13cm]{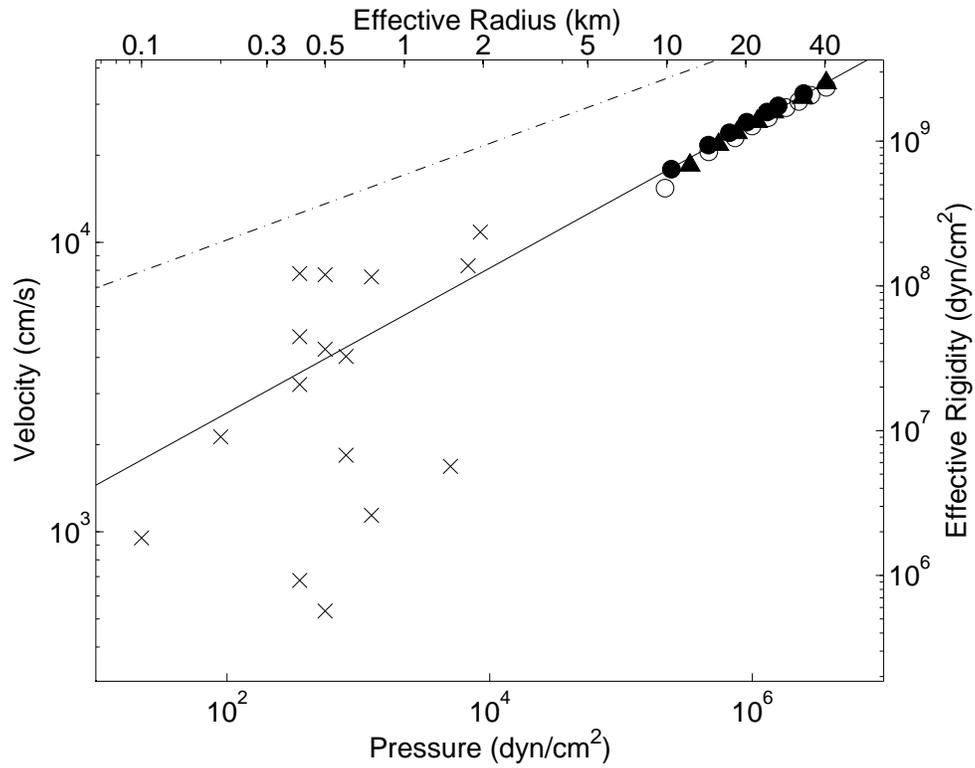}
\end{center}
\caption{Comparison of our model for the effective rigidity of
rubble piles with that from experiments on sand taken from
\cite{God90}. Shear wave velocity as a function of pressure in
saturated, dry, and drained Ottawa sands is shown by open circles,
solid circles, and triangles, respectively. Effective rigidities of
NEAs, inferred by assuming binary ages of 1Myr and $Q=100$, are
plotted against the primary diameter and marked by x's.}
\label{fig:experiment}
\end{figure*}

Next we compare data on the rigidity of sand with that on the
effective rigidity of NEAs. To do so, we assume that the semimajor
axes of binary NEAs have evolved from much smaller initial values
over $\sim 1\,$Myr with a tidal $Q=100$. Then we use equations
\refnew{tildemu}, \refnew{semimajor}, and \refnew{eq:k} to evaluate
the effective tidal rigidity of the primary for each of the binaries
in table \ref{table}. These rigidities are plotted as x's on figure
\ref{fig:experiment}. Although the scatter is large, probably
dominated by our assumption of a uniform age, the data fit nicely on
the extrapolation to low pressure of the data from the experiments
on sand.

\subsection{rates of eccentricity evolution in binary rubble piles}

Binary near-Earth asteroids are thought to form by
Yorp\footnote{Yorp spin up has been measured for near-Earth asteroid
2000 PH5 \citep{LFP+07,TMV+07}.} spin up and/or tidal disruption and
consequently be rubble piles \citep{WaR06b}. Most have nearly
circular orbits from which \cite{WaR06a} argue that tidal damping of
their orbital eccentricities proceeds 3 to 4 orders of magnitude
faster than would be expected for binary monoliths of comparable
size. A significant fraction of this increase must be due to the
reduced rigidity of a rubble pile as compared to a monolith since
$Q$ cannot be smaller than unity.

For rocky bodies, scaling from the tidal Love number of the Moon,
$k_{Moon}\approx 0.03$,
\begin{equation}
\tilde \mu \approx 1.5\times 10^8 \left(\km\over R\right)^2\, ,
\end{equation}
which corresponds to $\mu\approx 5\times 10^{11}\dyne\cm^{-2}$. Thus
from equation \refnew{eq:mintildemu} with $\epsilon_Y=10^{-2}$, we
obtain
\begin{equation}
\frac{\tilde\mu}{\tilde\mu_{rubble}}\lesssim \left(\tilde\mu
\epsilon_Y\right)^{1/2}\approx 10^3\frac{\km}{R}\, .
\label{eq:reducmueff}
\end{equation}
Since typical secondaries among near earth asteroid binaries have
radii of a few tenths of a kilometer, much if not all of the
increase in the inferred rates of eccentricity damping might be due
to an increase of $k$. However, it would not be surprising if a
contribution came from a reduction of $Q$.

We note that close encounters with Earth or other planets might
reset the eccentricities of binary NEAs on timescales comparable to
those at which they evolve under tides. This issue deserves
investigation.

\subsubsection{conditions for eccentricity damping in binary asteroids}.

If both primary and secondary were strength rather than gravity
dominated ({$\tilde\mu\gg 1$}), then the ratio of the rates of
eccentricity excitation and damping would be
\begin{equation}
{\cal R}={19 \over 28} \left( \rho_s \over \rho_p \right)^2 {R_s
\over R_p} {\tilde \mu_s \over \tilde \mu_p} {Q_s \over Q_p}\, .
\end{equation}
For monoliths of identical composition, this ratio reduces to
\begin{equation}
{\cal R}_{monolith}={19 \over 28} {R_p \over R_s}
{Q_s \over Q_p}\, .
\end{equation}
Thus for $Q_s/Q_p=1$,\footnote{Identical compositions do not
guarantee identical $Q$s, because the latter may also depend on
strain, strain rate, temperature, and pressure.} eccentricity
damping would require $R_p/R_s < 1.47$ corresponding to a mass ratio
less than 3.20. For primary and secondary composed of spherical
elements with identical compositions and $Q$'s, the ratio reads
\begin{equation}
{\cal R}_{spheres}={19 \over 28} \left( R_p \over R_s
\right)^{1/3}\, ,
\end{equation}
so eccentricity would damp for $R_p/R_s < 3.2$ corresponding to a
mass ratio below 33. Finally, for rubble piles composed of irregular
elements of identical compositions and $Q$'s,
\begin{equation}
{\cal R}_{rubble}={19 \over 28}
\end{equation}
so eccentricity would damp for all mass ratios.

It is clear that eccentricity damping is more likely for binary
rubble piles than for binary monoliths especially when the mass
ratio is not large. However, given the uncertainties in the relative
values of the primary's and secondary's $\tilde\mu_{rubble}$ and
$Q$, eccentricity growth remains a possibility, in particular for
large mass ratios.

\acknowledgments This research was supported in part by an NSF grant
and a NASA grant. RS is an Alfred P. Sloan Fellow, and a Packard
Fellow. We thank Hiroo Kanamori for valuable advice.


\end{document}